\documentclass[12pt, letterpaper]{article}
\usepackage[utf8]{inputenc}

\usepackage{subfigure}
\usepackage{graphicx}

\title{Sensitivity Improvements of Very-High-Energy $\gamma$-Ray Detection with the Upgraded H.E.S.S.~I Cameras using Full Waveform Processing}

\author{J.~Zorn\footnote{Max-Planck-Institut f\"ur Kernphysik, P.O. Box 103980, D 69029 Heidelberg, Germany, \texttt{justus.zorn@mpi-hd.mpg.de}} for the H.E.S.S.~collaboration\footnote{for collaboration list see PoS(ICRC2019)1177}}

\date{36th International Cosmic Ray Conference (ICRC2019)}

\begin{document}

\maketitle

\begin{abstract}
The High Energy Stereoscopic System (H.E.S.S.) is an array of five imaging atmospheric Cherenkov telescopes in Namibia observing $\gamma$-rays in the energy range from a few tens of GeV to a few tens of TeV. The Cherenkov signal detected by photomultiplier tubes is sampled at 1~GHz. In nominal data acquisition (charge) mode, this signal is integrated over a fixed window of 16~ns in case trigger conditions are met. Thanks to the electronics upgrade of the four H.E.S.S.~I cameras in spring 2017, full 1~GHz-sampled waveforms can be read out in parallel to the nominal charge mode. This allows for a higher flexibility in data analysis like signal integration along the signal time gradient, thereby increasing the signal-to-noise ratio and thus the sensitivity at the lower end of the energy range. Furthermore, it prevents the truncation of Cherenkov events lasting longer than 16~ns, enhancing the shower reconstruction of $\gamma$-ray events with TeV energies and high impact distances. Observations of PeVatron candidates may profit a lot from this new data acquisition mode since precise reconstruction of the rare multi-TeV $\gamma$-ray events is improved -- a crucial aspect to investigate a potential spectral cut-off.

Performance studies of the upgraded H.E.S.S.~I cameras with a focus on sample mode data analysis and comparison to nominal charge mode data are presented in this contribution.
\end{abstract}

\section{Introduction}
\label{sec:intro}
The High Energy Stereoscopic System (H.E.S.S.) array consists of five imaging atmospheric Cherenkov telescopes observing $\gamma$-rays in the energy range from a few tens of GeV to a few tens of TeV (cf.~e.g~\cite{Aha06}). In nominal data acquisition mode (referred to as charge mode, CM), the Cherenkov signal detected by the photomultiplier tubes (PMTs) is sampled at 1~GHz and integrated over a fixed window of 16~ns in case trigger conditions are met. The position of the integration window is anchored to the trigger time and is the same for all camera pixels.

\begin{figure}[tb]
\centering
\includegraphics[width=0.7 \textwidth]{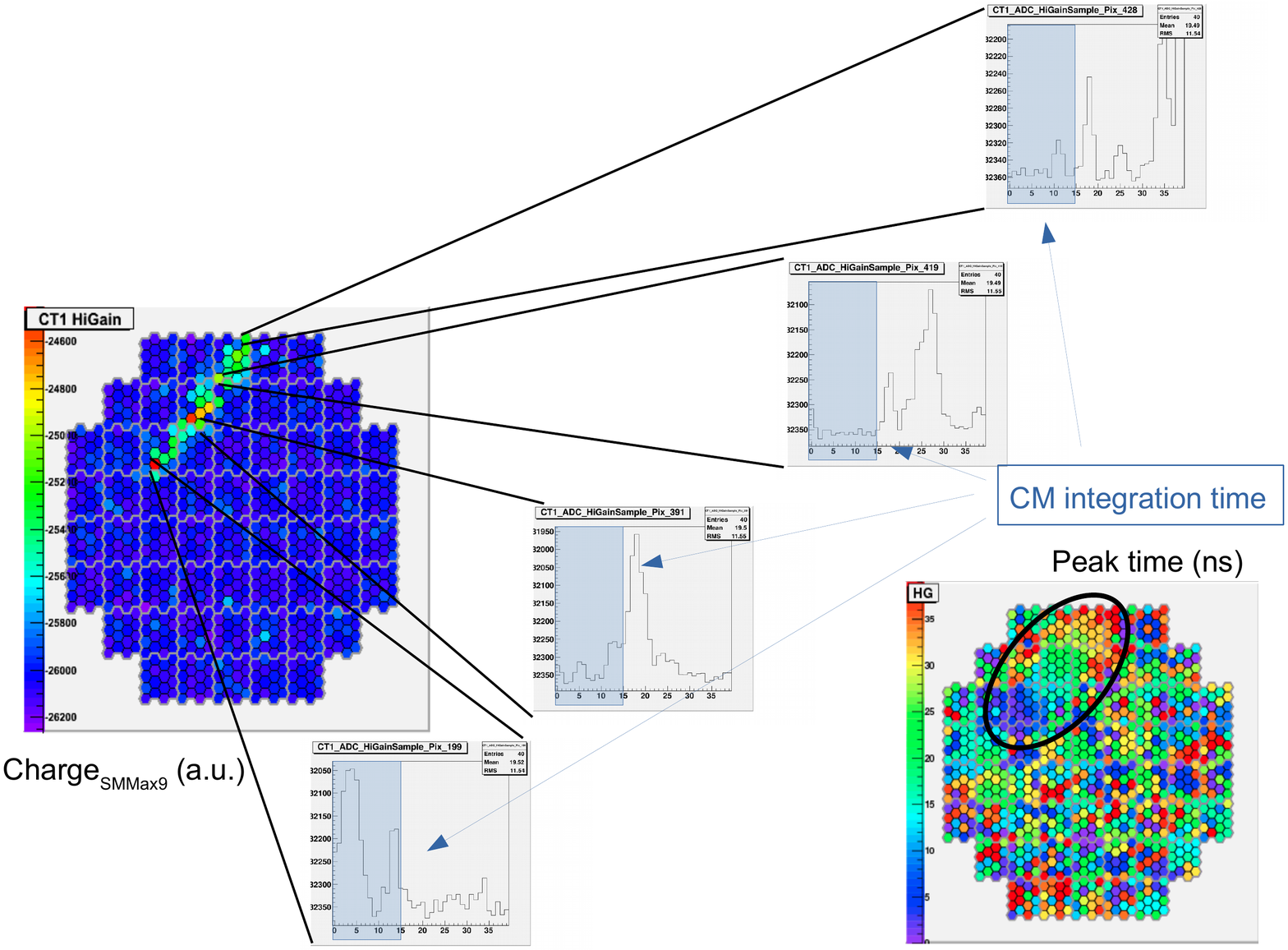}
\caption[]{Cherenkov images (raw data charge and peak time) are shown, taken in SM and using a fixed 9~ns-wide integration window around the peak maximum of each signal for charge extraction (referred to as SMMax9, cf.~Sec.~\ref{sec:extraction} for more details). Additionally, the corresponding uncalibrated (raw data) waveforms of a few pixels are displayed and the part of the signal (first 16~ns) being integrated in CM is indicated.}
\label{fig:image-truncation}
\end{figure}

\begin{figure}[tb]
\centering
\includegraphics[width=1 \textwidth]{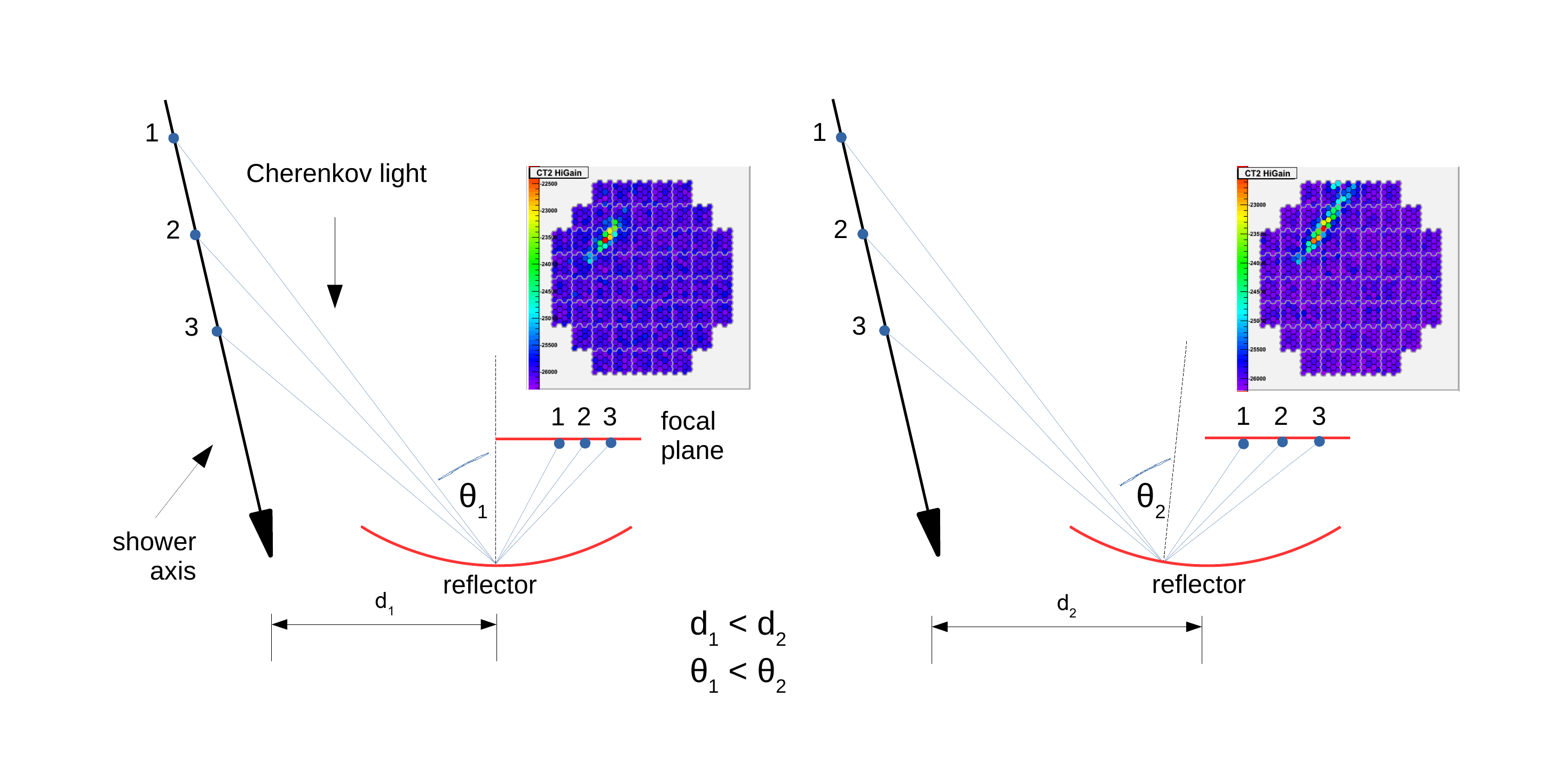}
\caption[]{Drawing showing the projection of Cherenkov light from extensive air showers into the camera of an imaging atmospheric Cherenkov telescope. Due to the geometric effect, Cherenkov events with high impact distances ($d_2$, right) typically produce longer images with a longer time gradient than those with small impact distances ($d_1$, left).}
\label{fig:iact-technique}
\end{figure}

In spring 2017, the four 14-year-old H.E.S.S.~I cameras (referred to as CT1--4) \cite{Vin03} underwent a major upgrade to improve their performance and robustness \cite{Gia17}\footnote{The upgraded cameras are referred to as HESS1U.}. All camera hardware but the PMTs (i.e.~hardware for trigger, readout, power, cooling and mechanical systems) were replaced. In combination with the use of new tailored software, this enabled the possibility to read out full 1~GHz-sampled waveforms with a maximum length of 40~ns in parallel to the nominal CM. This second data acquisition mode (referred to as sample mode, SM) allows a higher flexibility in data analysis like signal integration along the signal time gradient. It is expected to increase the signal-to-noise ratio and to prevent image truncation resulting in an increased sensitivity, energy and angular resolution at both, the lower and higher end of the energy range. Fig.~\ref{fig:image-truncation} is used to illustrate the latter effect (image truncation).
It shows a Cherenkov event with an image time spread exceeding the nominal 16~ns integration time. Hence it would appear truncated in CM reducing the reconstruction performance. Since events with a high impact distance are high-energy events (due to a higher light yield) and since the higher the impact distance of an event the longer the typical time gradient of the image (cf.~Fig.~\ref{fig:iact-technique}), the reconstruction performance of high-energy events will profit a lot from SM data taking.

Thanks to the readout in parallel to CM, the nominal data acquisition (DAQ) is preserved reducing the risk of incompatibility with pre-upgrade measurements and allowing a direct comparison between both modes.

Since SM data taking is not yet enabled by default in every observation run, two sources were specifically chosen to be observed in SM: Crab ($\sim$20 hours) mainly for technical verification and Westerlund I ($\sim$100 hours) mainly to expand the spectrum to higher energies to investigate a potential cutoff being a crucial criterion for the possible identification of Westerlund I as a PeVatron candidate.

These proceedings provide details about the readout scheme, the charge extraction algorithm, the incorporation into the existing H.E.S.S.~simulation, calibration, and analysis pipelines, and show performance results through the comparison of SM with CM.
\section{Readout scheme}
The current SM readout scheme 
is based on two aspects which are
\begin{enumerate}
\item separation from CM readout chain to reduce the risk of possible effects on data taken in CM and thus ensuring a stable and reliable nominal data taking while SM is introduced and validated,
\item readout and storage of full 40 samples-deep waveforms of all pixels without data reduction or other on-site operations to keep full flexibility for investigation.
\end{enumerate}
%
%
Due to the latter aspect -- the SM data being 40 times bigger in size (140~GB per camera for a 28~min run, assuming a mean trigger rate of 500~Hz) -- and the limited data transfer rate from the camera PCs to the DAQ process, the SM data is not transferred to the DAQ in real time in contrast to the CM data. Instead it is stored locally on the camera PCs. An offline event merger process, running after the observing shift (during daytime), builds the appropriate events from the runs stored on the different camera PCs. Consequently, CM and SM data are stored in two different \texttt{ROOT} \cite{ROOT96} files.

SM data readout is planed to be enabled in all observation runs in the near future. To do so, it will be reduced online and then transferred and processed in the same way as CM data via the DAQ, and stored as another data set in the same \texttt{ROOT} file. 
The reduction will be based on the same charge extraction mechanism as used for the performance studies (cf.~following section).
\section{Charge extraction algorithm and integration into analysis framework}
\label{sec:extraction}
To extract the SM charge from the waveform, the signal is integrated in a fixed window around the peak in each pixel. The so-called next neighbour (NN) peak finding algorithm is used to identify a Cherenkov signal in each pixel. It sums the waveforms of all pixels adjacent to the pixel of interest (PoI) and takes the time of the maximum value as peak time for the PoI. By default, the high-gain channel is used for this algorithm\footnote{The H.E.S.S.~cameras use two gain channels for processing the PMT signal to enlarge the dynamic range.}. In case the signal in one of the adjacent pixel exceeds the limit in which the signal response is linear, the low-gain channel is used for the NN peak finding of the PoI. Once the peak time of the PoI is found (same peak time used for high- and low-gain channel), the signal is integrated in a defined window around that peak resulting in one charge number per pixel for each gain channel. In the studies presented in these proceedings a window size of 9~ns was used with a shift to the left of the peak time of 3~ns. The extracted charge is referred to as ``SMMax9''.

The NN peak finding algorithm is an effective search for Cherenkov signals. Due to the use of the NNs and the timing properties of a Cherenkov signal, the mis-identification of an night-sky-background (NSB) photon as a Cherenkov photon is minimised, especially important for low-signal pixels. If for example just the time of the signal maximum in the PoI was taken as peak time, a mis-identification would happen very often as soon as the signal amplitude of Cherenkov and NSB photons are of the same order. Furthermore, using the sum of the NN waveforms leads to a higher weight of pixels with Cherenkov signal than those without. Another advantage of the algorithm is that no distinction between pixels with and without signals is needed. Thus, the same method can be used to extract the charges of no-signal pixels which are being used for the pedestal estimation further down in the calibration procedure. Due to the uncorrelated appearance of NSB signals in adjacent pixels, a summation of the waveforms smooths out any NSB signal. This prevents a bias to higher amplitudes in the integration which would lead to a higher (wrong) pedestal estimation.   

The integration of SM into the existing calibration chain is straight forward since it requires one additional step only -- the charge extraction -- preceding the nominal calibration chain. Once the charge has been extracted from the waveforms, the SM data can be processed in the same way as CM.

\begin{figure}[tb]
\centering
\includegraphics[width=0.7 \textwidth]{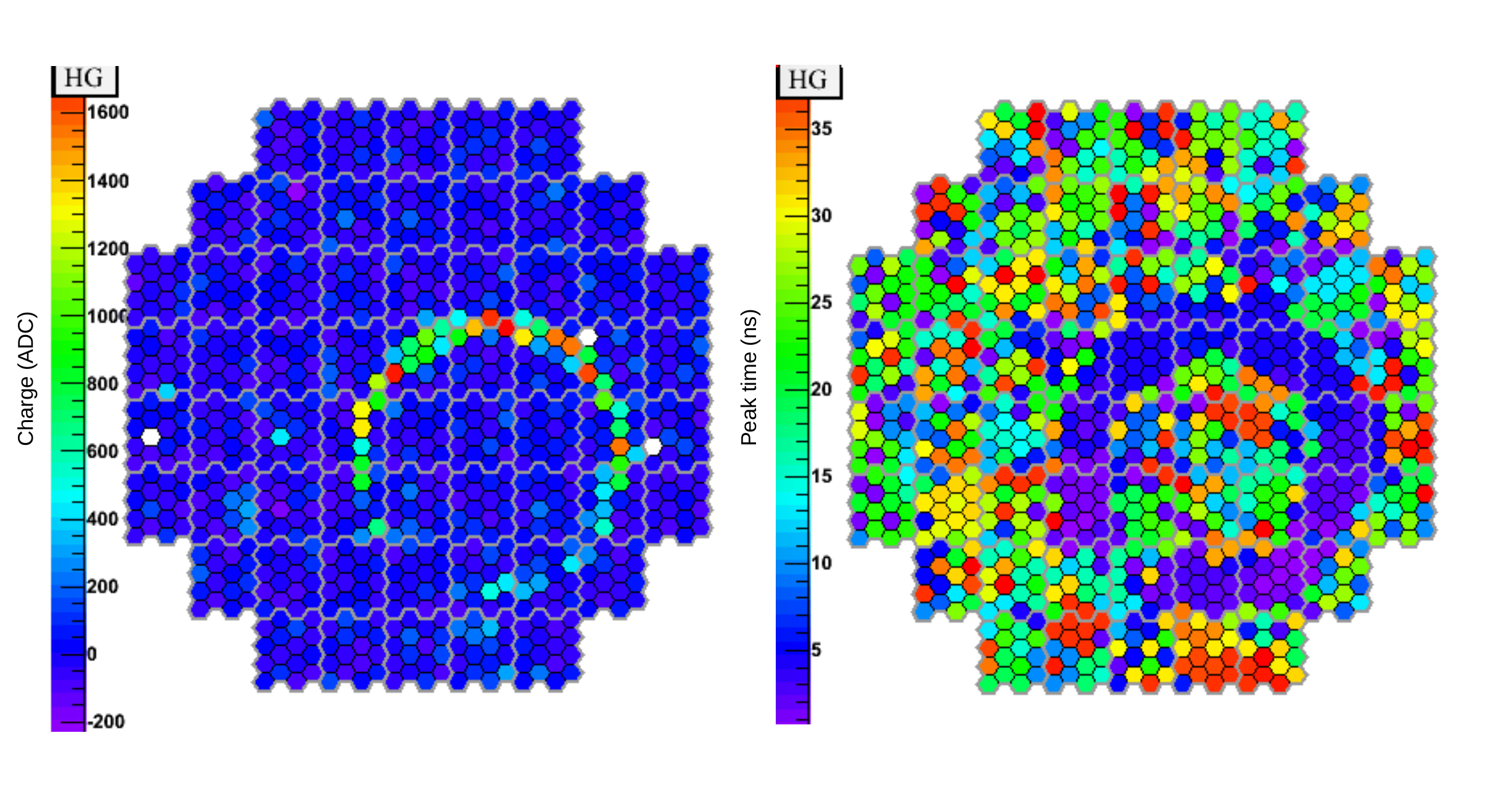}
\caption[]{Camera images of (left) extracted SMMax9 charge and (right) extracted peak time using the NN peak finding algorithm and the integration in a fixed 9~ns-wide window around the peak.}
\label{fig:muon-event-peaktime}
\end{figure}

The algorithm has been tested and validated on both, LED flasher runs (nominally used for flat-fielding) and observation runs. As an example, Fig.~\ref{fig:muon-event-peaktime} shows the extracted SMMax9 charge and peak time of a muon event. It illustrates the correct functionality of the NN peak finding algorithm.

\begin{figure}[tb]
\centering
\subfigure[]{\includegraphics[width=0.44 \textwidth]{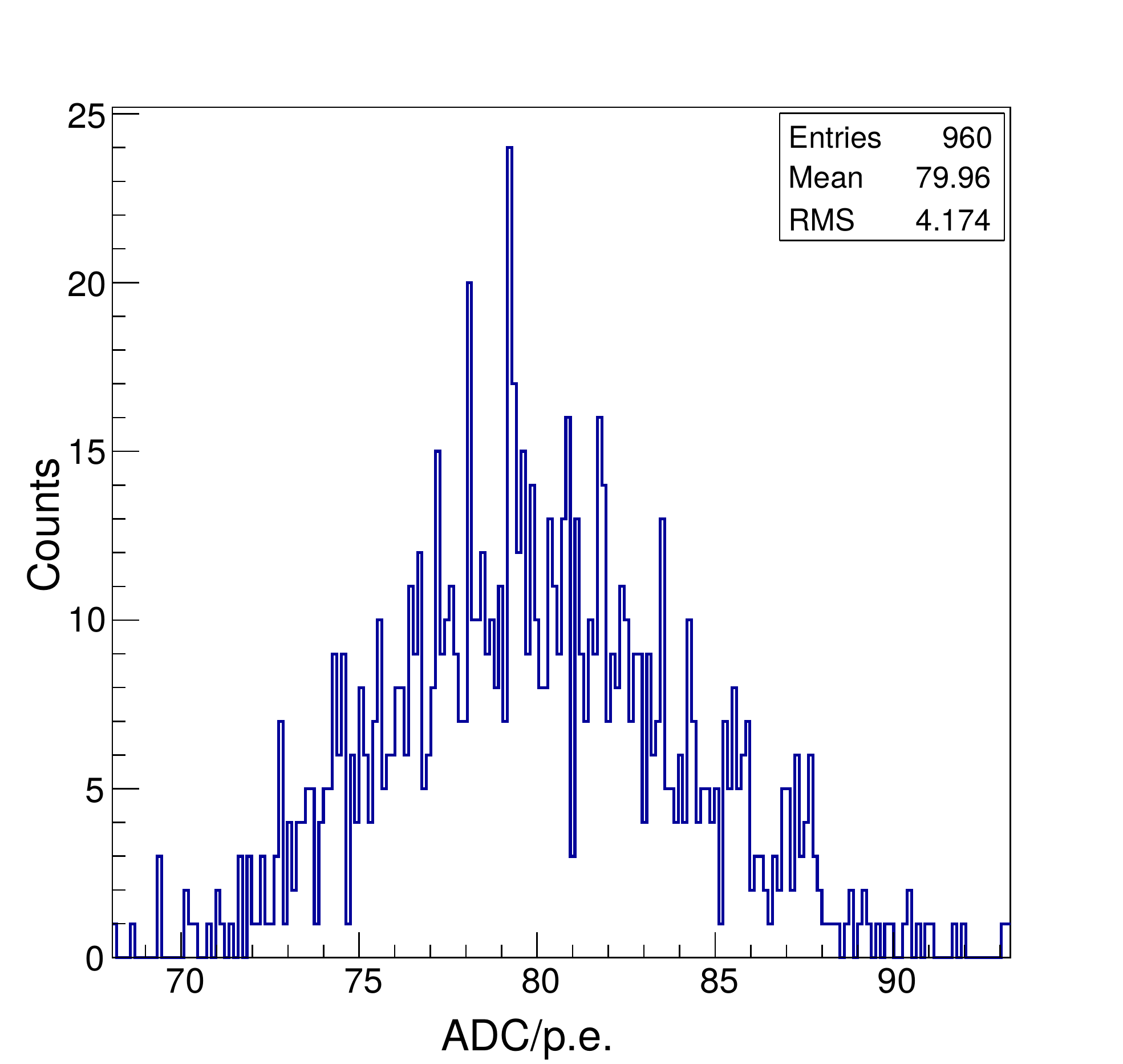}\label{fig:spe-cm}}
\subfigure[]{\includegraphics[width=0.44 \textwidth]{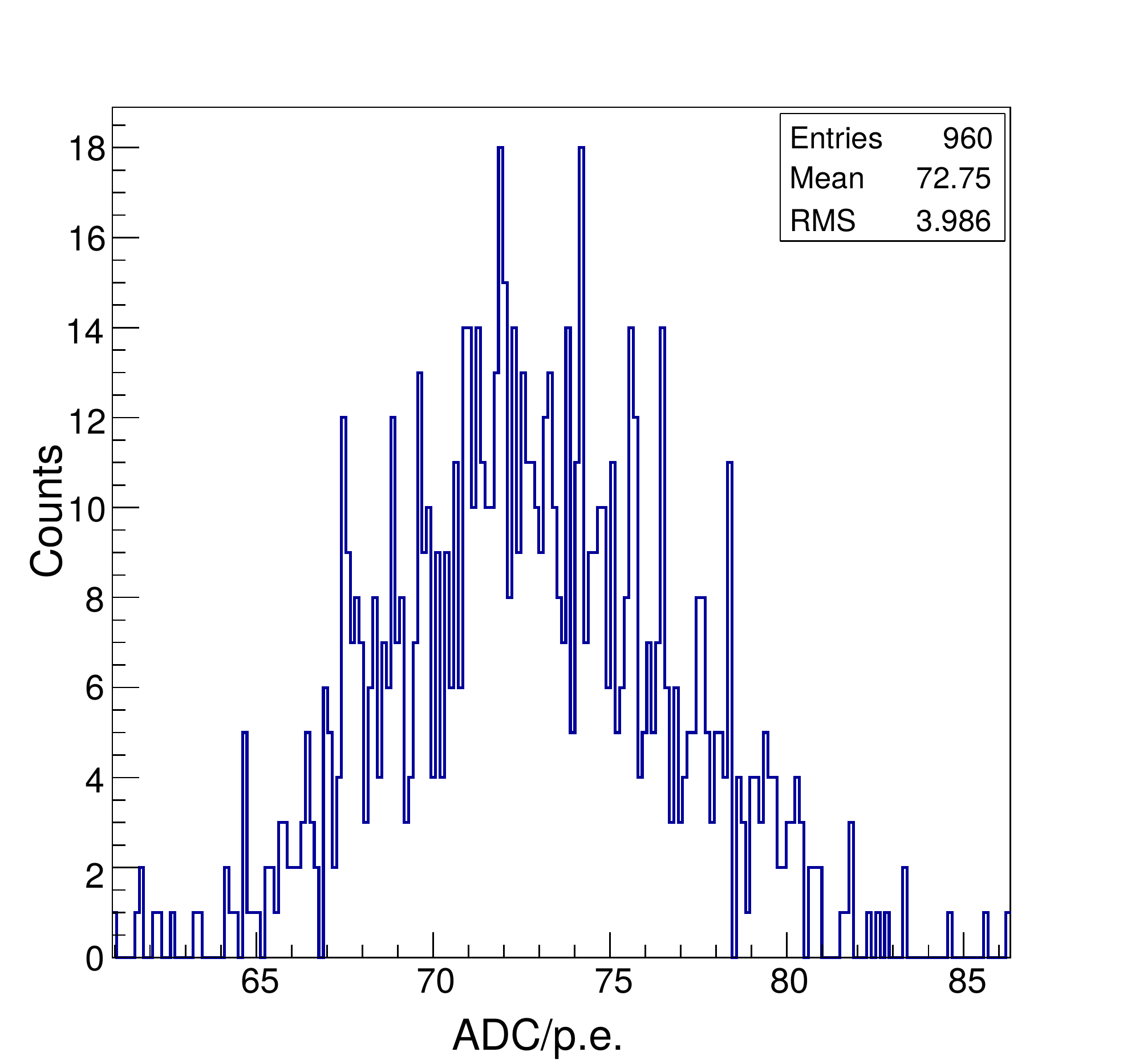}\label{fig:spe-sm}}
\caption[]{Gain distribution (in ADC/p.e.) for all 960 pixels of CT1 for (a) CM and (b) SM simulations.}
\label{fig:spe}
\end{figure}

\begin{figure}[tb]
\centering
\subfigure[]{\includegraphics[width=0.49 \textwidth]{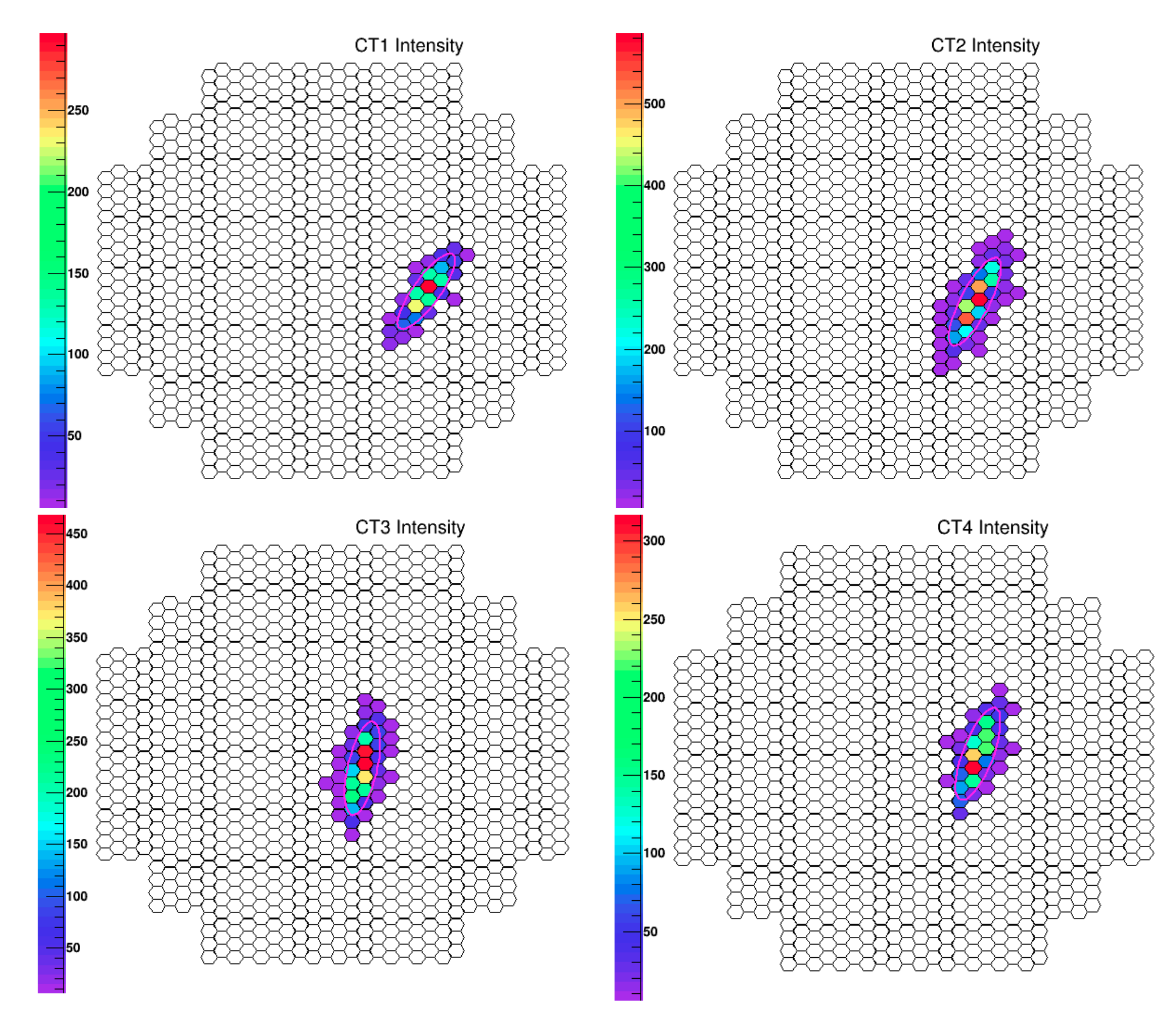}\label{fig:mc-cm}}
\subfigure[]{\includegraphics[width=0.49 \textwidth]{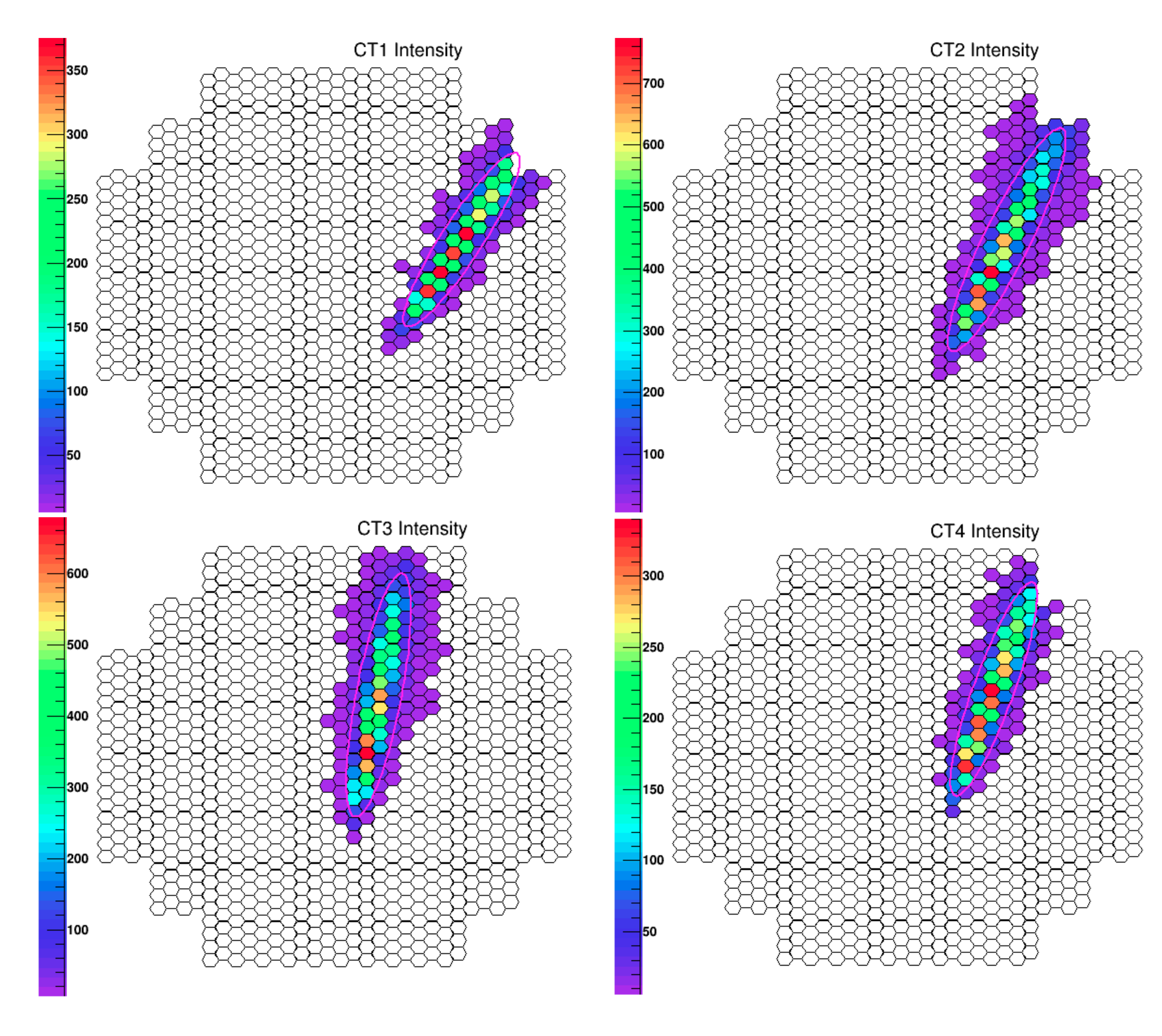}\label{fig:mc-sm}}
\caption[]{Simulated camera images showing the pixel-wise intensity in units of p.e.~(indicated by the colour scale) for (a) CM and (b) SMMax16. They result from an extensive air shower at a zenith angle of 45$^\circ$, being offset by 2$^\circ$ from the optical axis of the telescope, with an impact distance of 925~m, and initiated by a $\gamma$-ray with an energy of 223~TeV.}
\label{fig:MC-camera-image-CM-SMMax16}
\end{figure}

Similar to the calibration chain, SM has been incorporated into the simulation chain (based on \texttt{CORSIKA} \cite{1998cmcc.book.....H} and \texttt{sim\_telarray} \cite{Bernlohr:2008kv}). All HESS1U simulations are performed in SM by default. For CM, the first 16 samples are summed afterwards. For SM, the simulated samples are processed in the same way as for real data, i.e.~the same charge extraction based on peak finding and integration around peak maxima is applied. Fig.~\ref{fig:spe} shows the extracted gain in ADC/p.e.\footnote{p.e.~being the abbreviation for photoelectron} from single photoelectron (SPE) simulations revealing a factor of $\sim$0.91 between CM and SM gain.
This ratio is expected from the integration of the SPE pulse shape (being used as an input to the simulation chain) 
in two different windows (first 16~ns and 9~ns around the peak of the pulse shape).
%
%
Fig.~\ref{fig:MC-camera-image-CM-SMMax16} shows camera images of the same MC Cherenkov event resulting from a high-energy, high-impact-distance extensive air shower ($E \approx 223$~TeV, $d\approx 925$~m) in CM and SM. 
It nicely illustrates the truncation of CM images resulting in a different reconstruction of the Hillas ellipse with a significant lower length. Both, Fig.~\ref {fig:spe} \& \ref{fig:MC-camera-image-CM-SMMax16} serve as an example to show the correct incorporation of SM into the simulation chain.

For further processing of SM data in the analysis chain, different sets of cuts (like tail and local distance cuts as well as cuts on the image amplitude and $\gamma$-hadron-separation parameters) have to be defined. Furthermore, if based on a Hillas analysis, a new Hillas length-width-weighting needs to be used for the direction reconstruction. For the reconstruction based on \texttt{ImPACT} \cite{Par14}, new templates have to be produced. An optimisation study to define new cuts and new Hillas analysis settings, and to produce new \texttt{ImPACT} templates for SM data is underway.
\section{Performance studies}
\begin{figure}[tb]
\centering
\subfigure[]{\includegraphics[width=0.49 \textwidth]{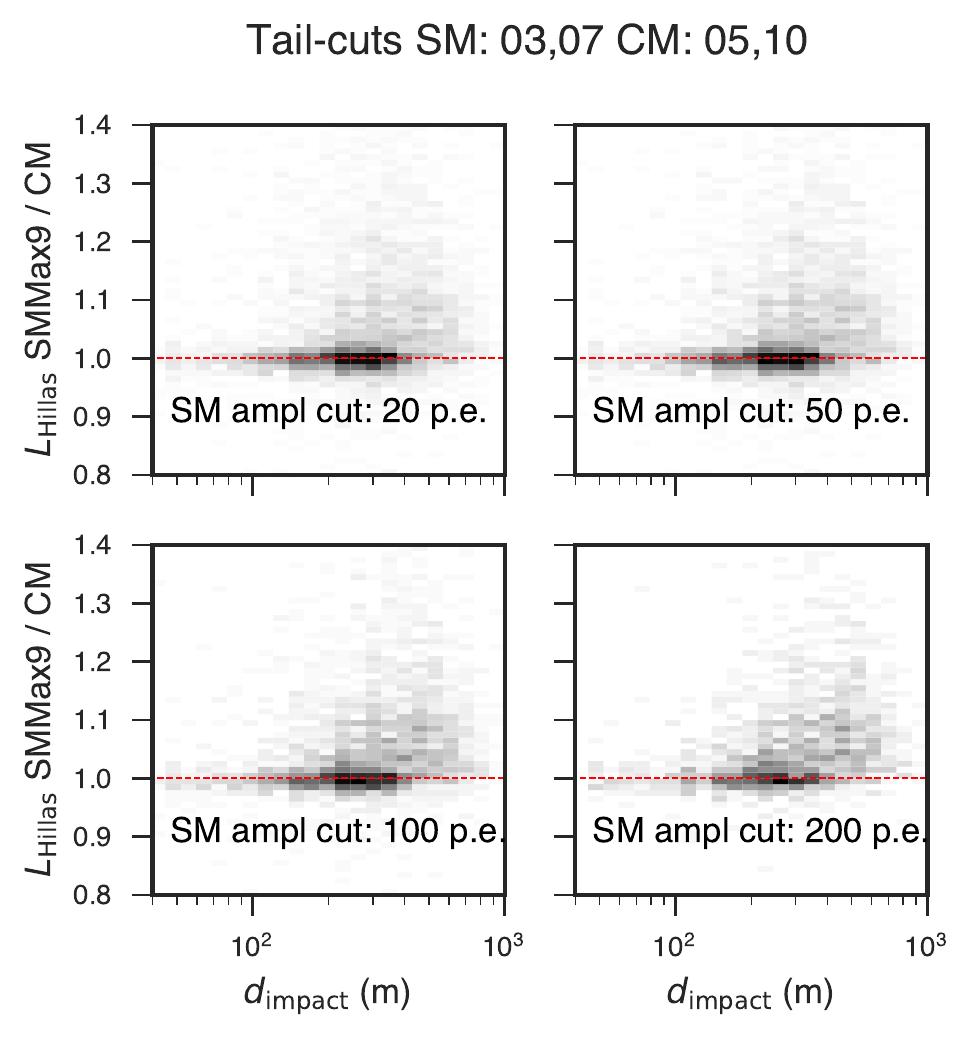}\label{fig:hillas-length-distance}}
\subfigure[]{\includegraphics[width=0.49 \textwidth]{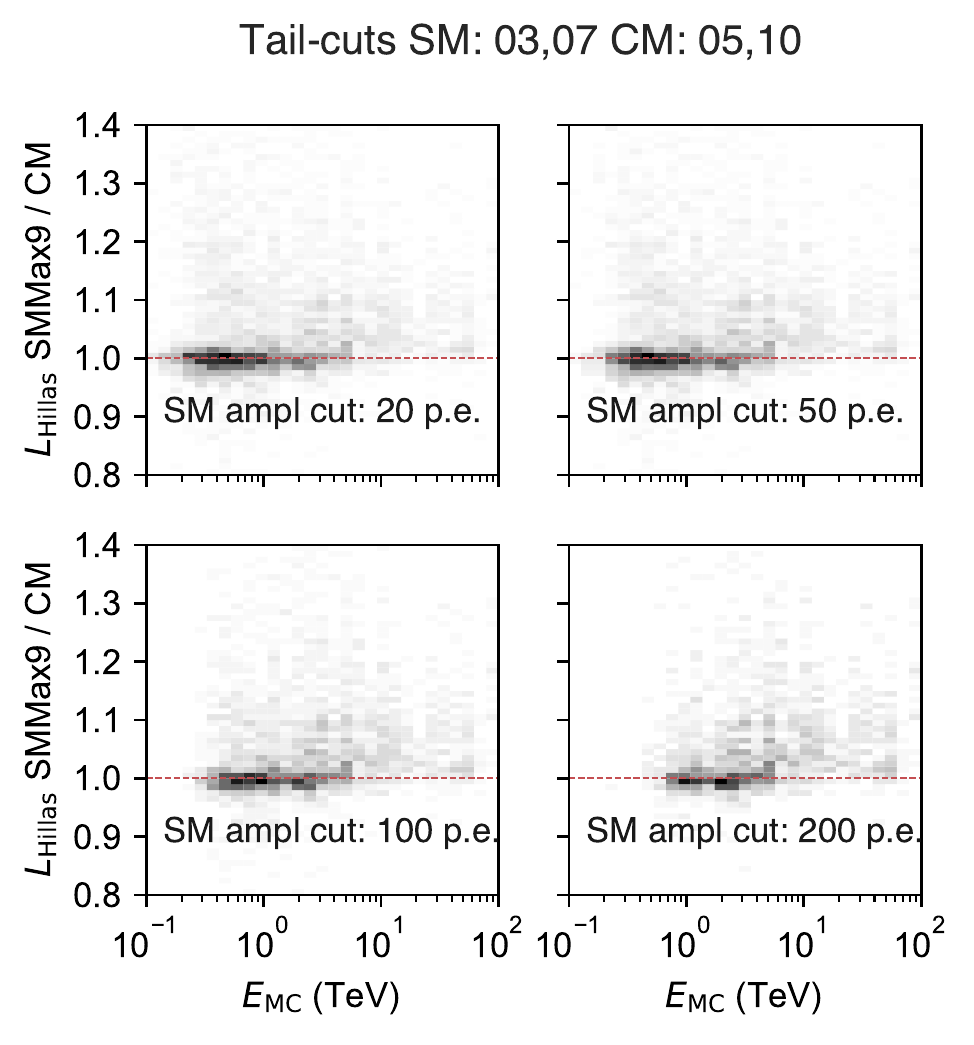}\label{fig:hillas-length-energy}}
\caption[]{Simulations of SMMax9 and CM Cherenkov events. Shown are 2D histograms with the ratio of SMMax9 and CM Hillas length $L_{\rm{Hillas}}$ as function of (a) impact distance $d_{\rm{impact}}$ and (b) energy of the primary $\gamma$-ray $E_{\rm{MC}}$ for different Hillas amplitude cuts applied (20, 50, 100, and 200~p.e.~. The number of counts in a bin is given by the colour scale (from white to black).}
\label{fig:hillas-length-comparisons}
\end{figure}

The performance of SM compared to CM can be tested on several levels. Fig.~\ref{fig:hillas-length-comparisons} shows the Hillas length -- being extracted from the parameterised Cherenkov camera image after image cleaning -- as function of the impact distance and MC energy of the primary $\gamma$-ray.
The image cleaning is based on a two-stage tail cut procedure to remove NSB fluctuations from the shower image. The tail-cut cleaning requires a pixel to have a signal greater than $x$ (or $y$) p.e.~and a NN pixel to have a signal larger than $y$ (or $x$)~p.e.~. For CM, $x$ and $y$ are nominally set to 5 and 10. For the integration of 9~ns instead of 16~ns, the noise is reduced by a factor $\sqrt{16/9}$. Furthermore, since the difference in gain (factor $\sim$0.91, cf.~Sec.~\ref{sec:extraction}) is not taken into account at the stage of the image cleaning, the values for tail cut cleaning should be reduced by a factor $\sqrt{16/9} \times 0.91$ for SM compared to CM. Hence, 3 and 7 are used for $x$ and $y$ in the SM image cleaning. As expected from the explanations given in Sec.~\ref{sec:intro} and from Fig.~\ref{fig:MC-camera-image-CM-SMMax16}, Cherenkov images recorded in SM are in general longer than those recorded in CM. This is mainly observed for events with high impact distances (cf.~Fig.~\ref{fig:hillas-length-distance}). Furthermore, Fig.~\ref{fig:hillas-length-energy} shows that not only the very bright high-energy events gain from being recorded in SM, also low-energy events may produce longer images in SM due to a reduced noise integration. There are also cases in which the Hillas length is slightly smaller in SM than in CM. In those cases, typically one more low-intensity pixel was cleaned away in SM than in CM resulting in a slightly reduced Hillas length.

\begin{figure}[tb]
\centering
\includegraphics[width=0.8 \textwidth]{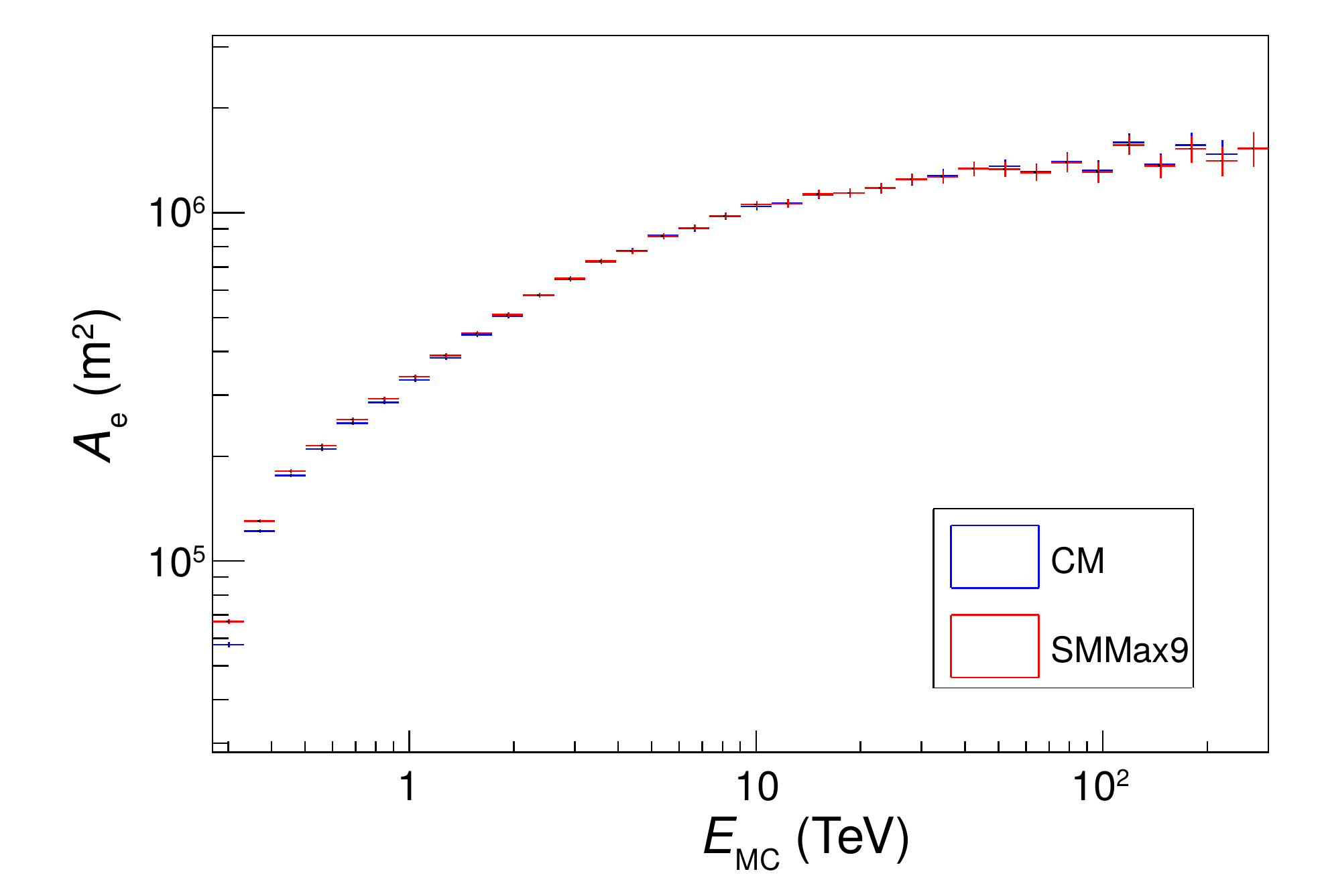}
\caption[]{Comparison of CM and SMMax9 effective areas $A_{\rm{e}}$ as function of MC energy $E_{\rm{MC}}$ with lower-image-amplitude cut applied only.}
\label{fig:comparison-effective-area}
\end{figure}

Once all cuts have been redefined for SM, the effective area is expected to improve at both the low- and high-energy end of the energy range since more images can be extracted in the analysis. Fig.~\ref{fig:comparison-effective-area} shows the effective area with a lower-image-amplitude cut applied only (60~p.e.~for CM, $\sim 0.91 \times 60$~p.e.~$\approx 54$~p.e.~for SM) and all other cuts being released.
As expected, the effective area is higher in SM at low energies due to the reduced noise integration. At high energies, it is similar for SM and CM what is expected since no further cuts are applied yet.

Studies of the difference in angular resolution between SM and CM is underway. First results indicate an improvement for SM, especially for events with high direction offsets ($\sim$2$^\circ$), from the optical axis of the telescope and primary $\gamma$-ray energies $>10$~TeV. In such cases, the images can be significantly longer ($\times 2$) in SM than in CM (cf.~Fig.~\ref{fig:MC-camera-image-CM-SMMax16}) improving the direction reconstruction performance and hence the angular resolution.
\section{Summary}
SM data taking with the HESS1U cameras has revealed to be reliable and to result in images with increased Hillas lengths and lower noise integration. The latter aspect was shown to result in increased statistics at the lower end of the H.E.S.S.~energy range. Same is expected at the higher end once the full set of new analysis cuts and new boosted decision trees (for $\gamma$-hadron separation) have been defined and re-trained for SM. Furthermore, a study about an improvement of energy and angular resolution is on-going. First results show an improvement of the latter one for data taken in SM, especially for high-offset and high-energy events.

So far, SM data was taken in two explicit observation campaigns only ($\sim$20 hours of Crab and $\sim$100 hours of Westerlund I observations). However, thanks to the studies presented in these proceedings promising better performance for all kinds of observations, it will be used by default in all observation runs in the near future. To be able to handle the increased amount of data, the SM data will be reduced online by using a charge extraction mechanism along the image time gradient. Thanks to the readout in parallel to CM, CM data is not affected and can further be used for standard analyses and backwards compatibility.
%
%

\end{document}